\documentclass{PoS}
\newcommand{\eq}[1]{Eq.~(\ref{#1})}
\title{Charged Higgs: Interpretation of $B$-physics results}
\usepackage{amsmath}
\usepackage{amssymb}
\usepackage{url}
\ShortTitle{2HDM flavour}

\author{\speaker{Andreas Crivellin}
\thanks{Work supported by a Marie Curie Intra-European Fellowship of the European Community's 7th Framework Programme under contract number (PIEF-GA-2012-326948).}\\
        CERN Theory Division,\\ CH-1211 Geneva 23,\\ SWITZERLAND\\
        E-mail: \email{andreas.crivellin@cern.ch}}

\abstract{In these proceedings we review the impact of additional Higgs bosons on $B$ physics observables. For this purpose, we consider first the 2HDM of type II which respects natural flavour conservation. Afterwards, we study a 2HDM with generic flavour structure (of type III). While constraints from $B_{s,d}\to\mu^+\mu^+$ impose stringent bounds on non-minimal flavour violation in the down sector, $b\to s,d\gamma$ can also constrain flavour violation in the up sector. Despite these stringent constraints and the recent CMS bounds from $A^0\to\tau^+\tau^-$ it is shown that still large effects in tauonic $B$ decays are possible, such that the BABAR measurements of these decays can be explained. The matching of the Yukawa sector of the MSSM on the 2HDM is discussed.}

\FullConference{Prospects for Charged Higgs Discovery at Colliders - Charged 2014,\\
		September 16-18, 2014,\\
		Uppsala University,\\ Sweden}

\begin{document}
\section{\label{sec:level1}Introduction}

The Standard Model (SM) possesses only one scalar $SU(2)_L$ doublet. After electroweak symmetry breaking, this gives masses to up quarks, down quarks and charged leptons. The charged (CP-odd) component of this doublet becomes the longitudinal component of the $W$ ($Z$) boson resulting in only one physical  Higgs particle. In a two-Higgs-doublet model (2HDM) \cite{Lee:1973iz} (see e.g. Ref.~\cite{Branco:2011iw} for a review) we introduce a second Higgs doublet and obtain four additional physical Higgs particles (in the case of a CP conserving Higgs potential): the neutral CP-even Higgs $H^0$, a neutral CP-odd Higgs $A^0$ and the two charged Higgses $H^{\pm}$. In the case of a MSSM-like Higgs potential we have $m_{A^0}\approx m_{H^0}\approx m_{A^0}\approx m_{H^\pm}\equiv m_{H}$ and $\tan\beta\approx-\cot\alpha$ for $v\ll m_{H}$ and $\tan\beta \gg 1$.

The most general Yukawa Lagrangian for quarks is given by
\begin{equation}
i \left( \Gamma_{q_f q_i }^{LR\, H} P_R + \Gamma_{q_f q_i }^{RL\, H} P_L \right) 
\end{equation}
with
\begin{eqnarray}
{\Gamma_{u_f u_i }^{LR\, H_k^0} } &=& x_u^k\left( \frac{m_{u_i }}{v_u}
\delta_{fi} - \epsilon_{fi}^{u}\cot\beta \right) + x_d^{k\star}
\epsilon_{fi}^{u}\,, \qquad
{\Gamma_{d_f d_i }^{LR\, H_k^0 } } = x_d^k \left( \frac{m_{d_i
}}{v_d} \delta_{fi} - \epsilon_{fi}^{d}\tan\beta \right) +
x_u^{k\star}\epsilon_{fi}^{ d} \,,\nonumber \\[0.1cm]
{\Gamma_{u_f d_i }^{LR\, H^\pm } } &=& \sum\limits_{j = 1}^3
{\sin\beta\, V_{fj} \left( \frac{m_{d_i }}{v_d} \delta_{ji}-
  \epsilon^{d}_{ji}\tan\beta \right)\, ,}\qquad
{\Gamma_{d_f u_i }^{LR\,H^ \pm } } = \sum\limits_{j = 1}^3
{\cos\beta\, V_{jf}^{\star} \left( \frac{m_{u_i }}{v_u} \delta_{ji}-
  \epsilon^{u}_{ji}\tan\beta \right)\, }\, .\nonumber
 \label{Higgs-vertices-decoupling}
\end{eqnarray}
Here, $H^0_k=(H^0,h^0,A^0)$ and the coefficients $x_q^{k}$ are given by
\begin{equation}
x_u^k \, = \, \frac{1}{\sqrt{2}}\left(-\sin\alpha,\,-\cos\alpha,\,i\cos\beta\right) \,,\qquad
x_d^k \, = \,\frac{1}{\sqrt{2}}\left(-\cos\alpha,\,\sin\alpha,\,i\sin\beta\right) \, .
\end{equation}
\smallskip
The ``non-holomorphic" couplings $\epsilon^q_{ij}$ parametrize the couplings of up (down) quarks to the down (up) type Higgs doublet\footnote{Here the expression  already implicitly refers to the MSSM where non-holomorphic couplings involving the complex conjugate of a Higgs field are forbidden.}. In the MSSM at tree-level we have $\epsilon^q_{ij}=0$. This corresponds to the 2HDM of type II which respects natural flavour conservation resulting in the absence of flavour changing neutral Higgs couplings. However, at the loop-level, the non-holomorphic couplings $\epsilon^q_{ij}$ are generated~(see for example Ref~\cite{Hamzaoui:1998nu}).

Tauonic $B$-meson decays are an excellent probe of charged Higgs bosons due to the heavy $\tau$ lepton involved. Recently, the BABAR collaboration performed an analysis of the semileptonic $B$ decays $B\to D\tau\nu$ and $B\to D^*\tau\nu$ using the full available data set~\cite{BaBar:2012xj}. They found for the ratios
\begin{equation}
{\cal R}(D^{(*)})\,=\,{\cal B}(B\to D^{(*)} \tau \nu)/{\cal B}(B\to D^{(*)} \ell \nu)\,,
\end{equation}
the following experimental values:
\begin{eqnarray}
{\cal R}(D)\,=\,0.440\pm0.058\pm0.042  \,,\qquad {\cal R}(D^*)\,=\,0.332\pm0.024\pm0.018\,.
\end{eqnarray}
Here the first error is statistical and the second one is systematical. If one compares these measurements to the SM predictions
\begin{eqnarray}
{\cal R}_{\rm SM}(D)\,=\,0.297\pm0.017 \,, \qquad{\cal R}_{\rm SM}(D^*) \,=\,0.252\pm0.003 \,,
\end{eqnarray}
one sees that there is a discrepancy of 2.2\,$\sigma$ for $\cal{R}(D)$ and 2.7\,$\sigma$ for $\cal{R}(D^*)$ and combining them gives a $3.4\, \sigma$ deviation from the SM~\cite{BaBar:2012xj}. This evidence for NP is also supported by the measurement of $B\to \tau\nu$ 
\begin{equation}
{\cal B}[B\to \tau\nu]=(1.15\pm0.23)\times 10^{-4}\,,
\end{equation}
which is by $1.6\, \sigma$ slightly higher than the SM prediction using $V_{ub}$ from a global fit of the CKM matrix \cite{Charles:2004jd}. A charged Higgs affects $B\to \tau\nu$~\cite{Hou:1992sy}, $B\to D\tau\nu$~\cite{Tanaka:1994ay} and $B\to D^*\tau\nu$\cite{Fajfer:2012vx}.

\section{The 2HDM of type II}

\begin{figure}[ht]
\begin{center}
\includegraphics[width=0.49\textwidth]{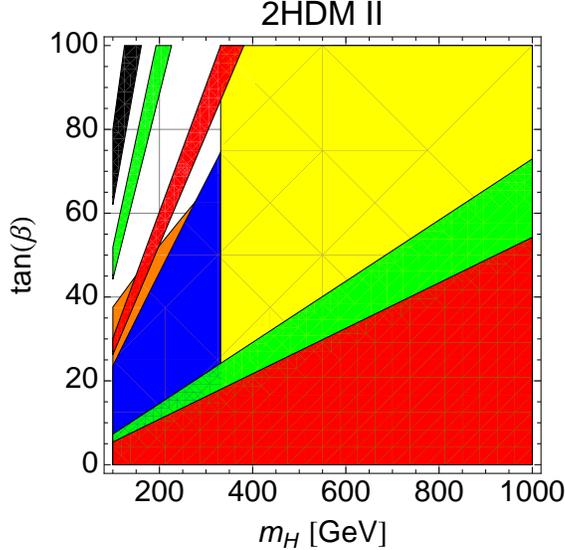}
\end{center}
\caption{Updated constraints on the 2HDM of type II parameter space from flavour observables \cite{Crivellin:2013wna}. The regions compatible with experiment are shown (the regions are superimposed on each other): $b\to s\gamma$~\cite{Hermann:2012fc} (yellow), $B\to D\tau\nu$ (green), $B\to \tau \nu$ (red), $B_{s}\to \mu^{+} \mu^{-}$ (orange), $K\to \mu \nu/\pi\to \mu \nu$ (blue) and $B\to D^*\tau \nu$ (black). Note that no region in parameter space is compatible with all processes. The tension originates from the destructive interference of the Higgs contribution with the SM one in $B\to D^*\tau \nu$ which would require very small Higgs masses and large values of $\tan\beta$ not compatible with the other observables. To obtain this plot, we added the theoretical uncertainty linearly on top of the $2 \, \sigma$ experimental error.}
\label{fig:2HDMII}
\end{figure}

As described in the last section, in the 2HDM of type II all flavour violation is due to the CKM matrix and we have (assuming a MSSM-like Higgs potential) only two additional free parameters with respect to the SM: $\tan\beta$ and the heavy Higgs mass $m_H$. In Fig.~\ref{fig:2HDMII} we show the allowed region in the $m_H$--$\tan\beta$ plane from various flavour observables~\cite{Crivellin:2013wna} (see also e.g. \cite{Deschamps:2009rh}). As we can see, for lower values of $\tan\beta$ ($\leq20$) and Higgs masses above 380~GeV all constraints except $B\to D^*\tau\nu$ can be satisfied within $2\sigma$ uncertainties. Due to the destructive interference of the SM with the charged Higgs contribution, very large values of $\tan\beta$ and very light Higgs masses would be required to explain $B\to D^*\tau\nu$. Such points in parameter space are not only ruled out by the other flavour observables but also by the $A^0\to\tau^+\tau^-$ bounds~\cite{CMS:2013hja}. Therefore, a 2HDM of type II cannot explain the observed anomalies in tauonic $B$ decays\cite{BaBar:2012xj}.

\section{The 2HDM of type III}

\begin{figure}[t]
\centering
\includegraphics[width=0.4\textwidth]{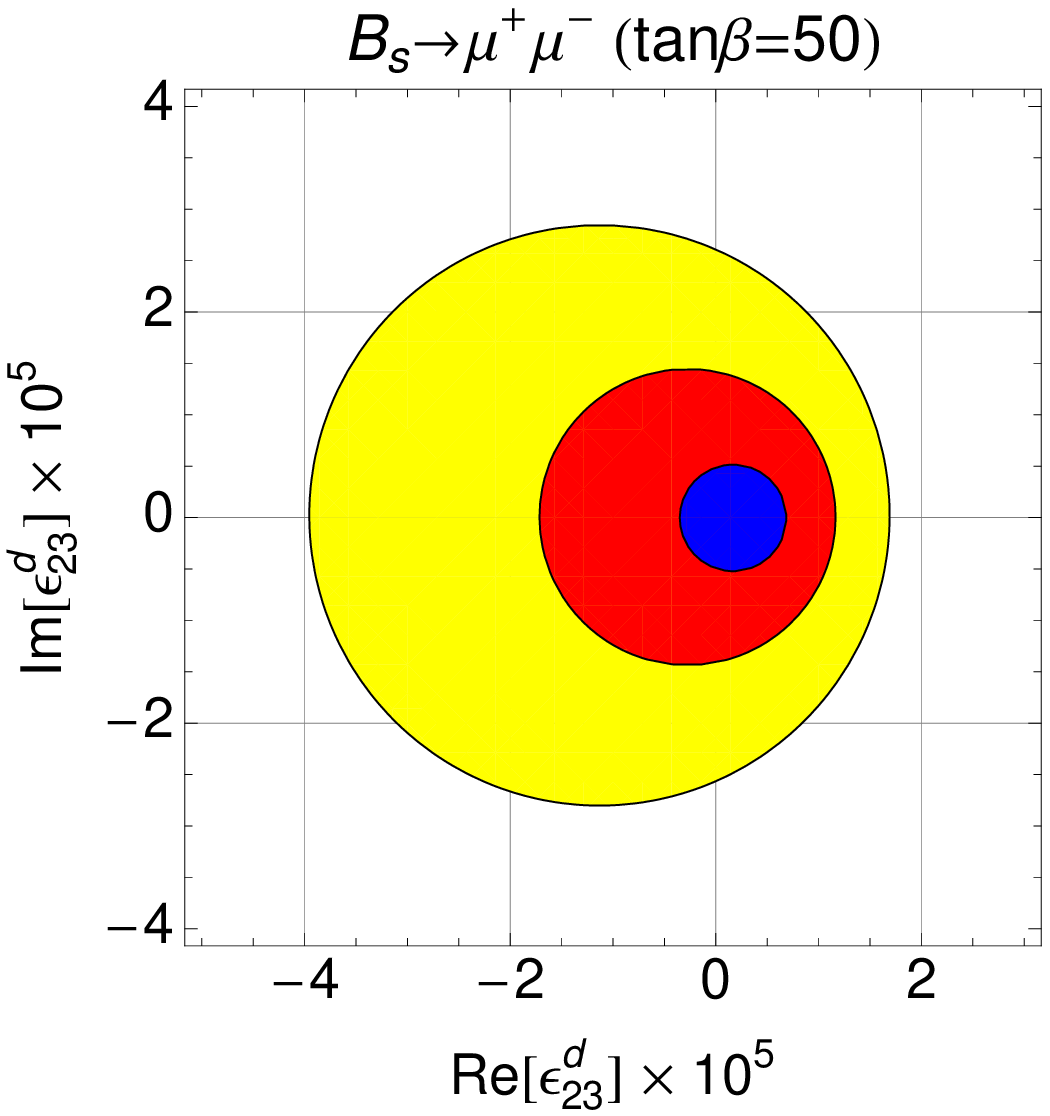}~~~
\includegraphics[width=0.444\textwidth]{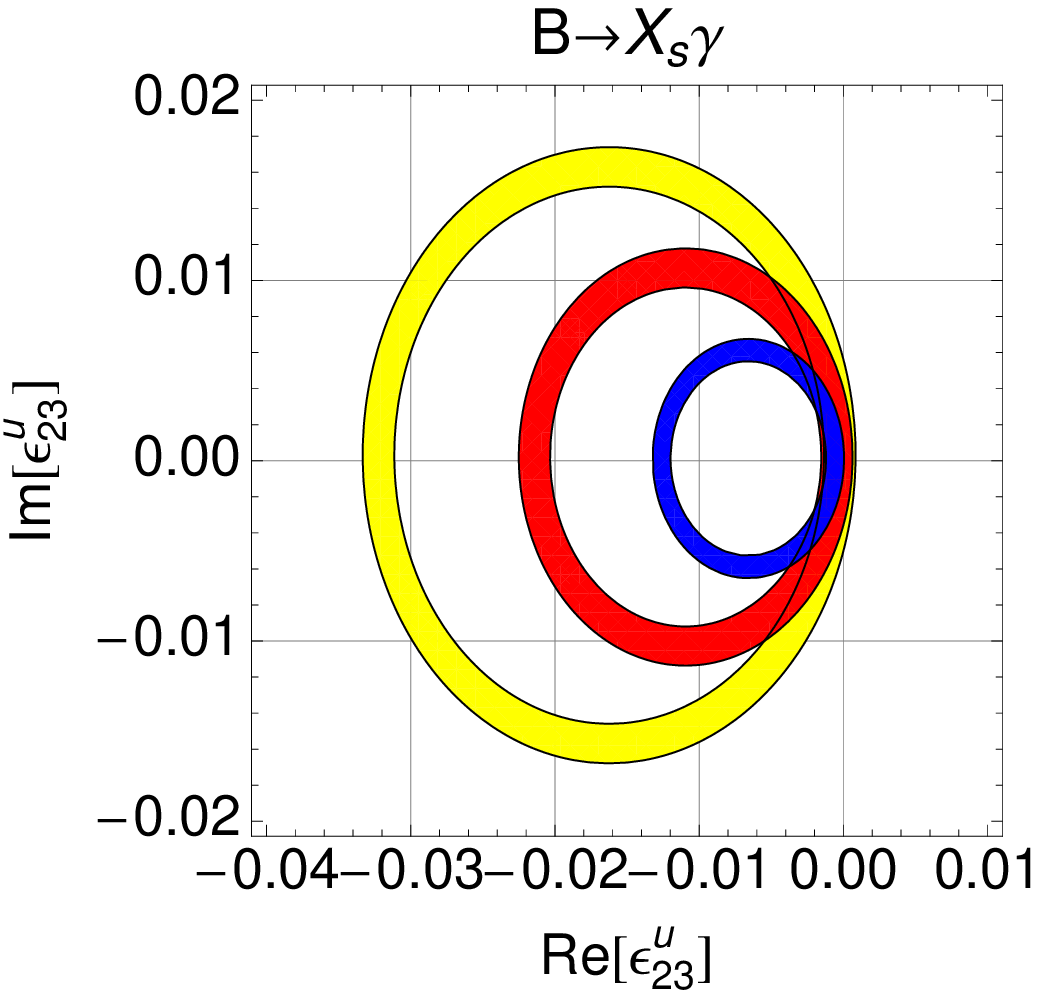}
\caption{Left: Allowed regions in the complex $\epsilon^{d}_{23}$--plane from $B_s\to\mu^+\mu^-$ for $\tan\beta=30$ and $m_{H}=700\mathrm{~GeV}$ (yellow), $m_{H}=500\mathrm{~GeV}$ (red) and $m_{H}=300\mathrm{~GeV}$ (blue). Note that the allowed regions for $\epsilon^{d}_{32}$--plane are not full circles because in this case a suppression of ${\cal B}\left[B_{s}\to\mu^+\mu^-\right]$ below the experimental lower bound is possible.\newline
Right: Allowed regions for $\epsilon^{u}_{23}$ from $ B \to X_{s} \gamma$, obtained by adding the $2\,\sigma$ experimental error and theoretical uncertainty linear for $\tan\beta=50$ and $m_{H}=700 \, \mathrm{ GeV}$ (yellow), $m_{H}=500\, \mathrm{ GeV}$ (red) and  $m_{H}=300 \,\mathrm{ GeV}$ (blue). \label{fig:Bstomumu}}
\end{figure}

In the 2HDM with generic flavour structure (i.e. of type III) the parameters $\epsilon^q_{ij}$ generate flavour chaining neutral Higgs couplings\cite{Sher:1991km}. Direct constraints on the off-diagonal elements $\epsilon^q_{fi}$ can be obtained from neutral Higgs contributions to the leptonic neutral meson decays ($B_{s,d}\to\mu^+\mu^-$, $K_L\to\mu^+\mu^-$ and ${\bar D}^0\to\mu^+\mu^-$) which arise already at the tree level\footnote{In principle, the constraints from these processes could be weakened, or even avoided, if $\epsilon^\ell_{22}\approx m_{\ell_{2}}/v_u$. Anyway, in here we will assume that the Peccei-Quinn breaking for the leptons is small and neglect the effect of $\epsilon^\ell_{22}$ in our numerical analysis for setting limits on $\epsilon^q_{ij}$.}. $K_L\to\mu^+\mu^-$ constrains $\left|\epsilon^d_{12,21}\right|$, $D^0\to\mu^+\mu^-$ imposes bounds on $\left|\epsilon^u_{12,21}\right|$ and $B_s\to\mu^+\mu^-$ ($B_d\to\mu^+\mu^-$) limits the possible size of $\left|\epsilon^d_{23,32}\right|$ $\left(\left|\epsilon^d_{13,31}\right|\right)$. We find the following (approximate) bounds on the absolute value of $\epsilon^q_{ij}$:
\begin{equation}
\renewcommand{\arraystretch}{1.4}
\begin{array}{l}
\left|\epsilon^d_{12,21}\right|\leq 6.4\times 10^{-6}\,,\qquad
\left|\epsilon^u_{12,21}\right|\leq 1.2 \times 10^{-1}\,,\\
\left|\epsilon^d_{23,32}\right|\leq 1.2 \times 10^{-4}\,,\qquad
\left|\epsilon^d_{13,31}\right|\leq 4.0 \times 10^{-5}\,,\\
\end{array}
\end{equation}
for $\tan\beta=50$ and $m_H=500$~GeV. As an example, we show the full dependence of the constraints in the complex $\epsilon^{d}_{23}$-plane from $B_s\to\mu^+\mu^-$ in left plot of Fig.~\ref{fig:Bstomumu}. Note that both an enhancement or a suppression of ${\cal B}\left[ B_{d,s}\to\mu^+\mu^-\right]$ compared to the SM values is possible. If at the same time both elements $\epsilon^{d}_{23}$ and $\epsilon^{d}_{32}$ are non-zero, constraints from $B_s$ mixing arise which are even more stringent.
\medskip

So far we were able to constrain all flavour off-diagonal elements $\epsilon^d_{ij}$ and $\epsilon^u_{12,21}$ but no relevant tree-level constraints on $\epsilon^u_{13,31}$ and $\epsilon^u_{23,32}$ can be obtained due to insufficient experimental data for top FCNCs. Nonetheless, it turns out that also the elements $\epsilon^u_{13,23}$ can be constrained from charged Higgs contributions to the radiative $B$ decay $b\to d \gamma$ and $ b\to s \gamma$. As an example we show the constraints on $\epsilon^u_{23}$ in the right plot of Fig.~\ref{fig:Bstomumu}. The constraints on $\epsilon^u_{13}$ from $ b\to d \gamma$ are even more stringent~\cite{Crivellin:2011ba}.

However, there are no relevant constraints on $\epsilon^u_{32,31}$ from FCNC processes because the light charm or up quark propagating in the loop contributes proportionally to this small mass. This has significant implications for charged current processes, i.e. tauonic $B$ decays.

\subsection{Tauonic $B$ decays}

\begin{figure}[ht]
\centering
\includegraphics[width=0.4\textwidth]{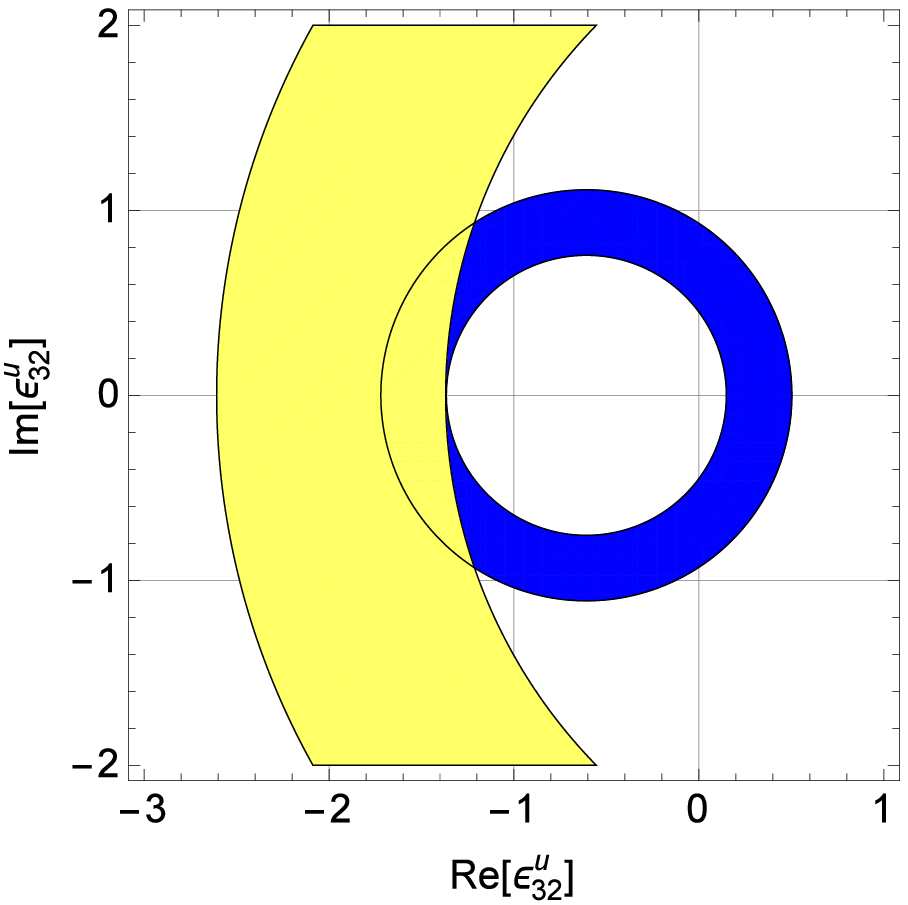}
\hspace{5mm}
\includegraphics[width=0.45\textwidth]{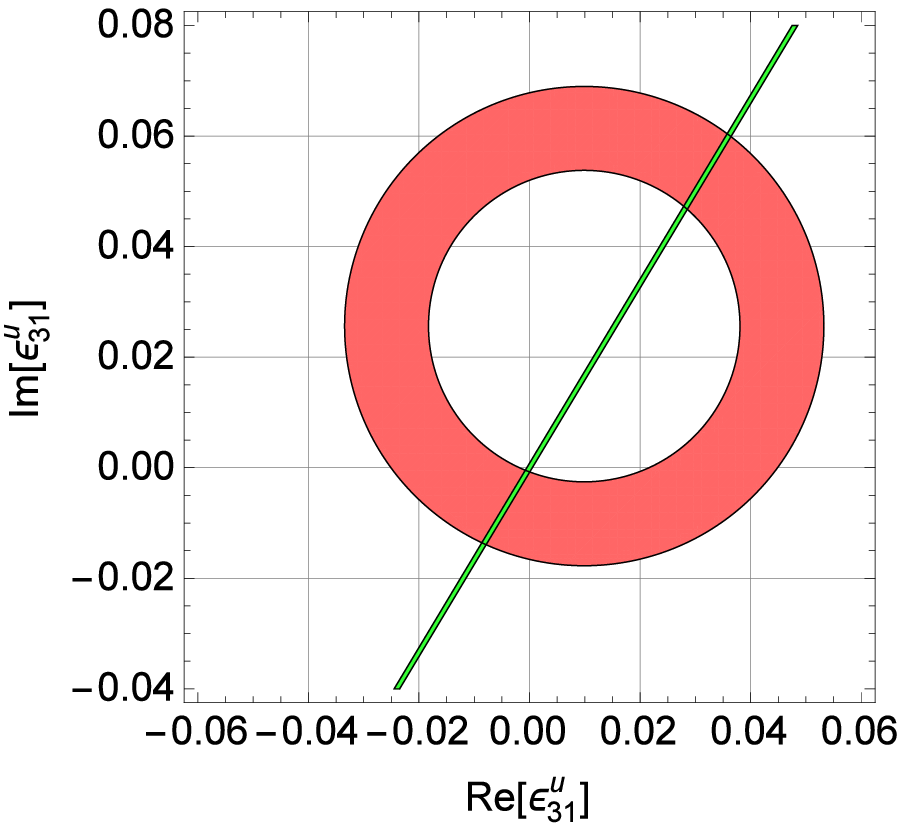}
\caption{Left: Allowed regions in the complex $\epsilon^u_{32}$-plane from $\cal{R}(D)$ (blue) and $\cal{R}(D^*)$ (yellow) for $\tan\beta=40$ and $m_H=800$~GeV. Right:  Allowed regions in the complex $\epsilon^u_{31}$-plane from $B\to \tau\nu$ (red) and the neutron EDM (green). \label{2HDMIII}}
\end{figure}

We found that all parameters $\epsilon^d_{ij}$ are stringently constrained from FCNC processes in the down sector. Furthermore, since $\epsilon^u_{23}$ ($\epsilon^u_{13}$) are constrained from $b\to s\gamma$ ($b\to d\gamma$) only $\epsilon^u_{31}$ ($\epsilon^u_{32}$) can significantly effect $B\to \tau\nu$ ($\cal{R}(D)$ and $\cal{R}(D^*)$). This even happens without any suppression by small CKM elements. Anyway, the lower bounds on the charged Higgs mass for a 2HDM of type II from $b\to s\gamma$ of 380~GeV \cite{Hermann:2012fc} must still be respected by our model (unless $\epsilon^u_{23}$ generates a destructively interfering contribution), and also the results from direct searches at the LHC for $A^0\to\tau^+\tau^-$~\cite{CMS:2013hja} are not significantly affected (unless $\epsilon^\ell_{33}$ is very large). 

Indeed, it turns out that by using $\epsilon^u_{32,31}$ we can explain $\cal{R}(D^*)$ and $\cal{R}(D)$ simultaneously~\cite{Crivellin:2012ye} (see also \cite{Celis:2012dk}). In Fig.~\ref{2HDMIII} we see the allowed region in the complex $\epsilon^u_{32}$-plane, which gives the correct values for $\cal{R}(D)$ and $\cal{R}(D^*)$ within the $1\, \sigma$ uncertainties for $\tan\beta=40$ and $M_H=800$~GeV. Similarly, $B\to \tau\nu$ can be explained by using $\epsilon^u_{31}$ while in this case the neutron EDM imposes stringent bounds on the phase of $\epsilon^u_{31}$\footnote{Note that $B\to\tau\nu$ might also be affected by a right-handed $W$ coupling \cite{Crivellin:2009sd}.}. These regions in parameter space are compatible with current CMS $A^0\to\tau^+\tau^-$~\cite{CMS:2013hja} bounds.

\section{Effective Higgs Vertices in the MSSM}
\label{sec:MSSM-2HDM}

In this section we discuss the matching of the MSSM on the 2HDM considering the Yukawa sector. As mentioned in the introduction, at tree-level the MSSM is a 2HDM of type II but at the loop-level, the Peccei Quinn symmetry of the Yukawa sector is broken by terms proportional to the higgsino mass parameter $\mu$ (or non-holomorphic $A^\prime$ terms). 

In the MSSM there is a one-to-one correspondence between Higgs-quark-quark couplings and chirality changing quark self-energies (in the decoupling limit\footnote{The non-decoupling corrections are found to be very small \cite{Crivellin:2010er}.}): The Higgs-quark-quark coupling can be obtained by dividing the expression for the self-energy by the vev of the corresponding Higgs field. 

Let us denote the contribution of the quark self-energy with squarks and gluinos to the operator $\overline{q}_f P_R q_i$ by $C_{f i }^{q\,LR}$. It is important to note that this Wilson coefficient is linear in $\Delta^{q\,LR}$, the off-diagonal element of the squark mass matrix connecting left-handed and right-handed squarks. For down squarks we have
\begin{equation}
	\Delta^{d\,LR}_{ij}=-v_d A^d_{ij}-v_u \mu Y^{d_i}\delta_{ij}\,,
\end{equation}
where the term $v_d A^d_{ij}$ originates from a coupling to $H^d$ while the term $v_u \mu Y^{d_i}$ stems from a coupling to $H^u$ (and similarly for up-squarks). Thus, we denote the piece of $\hat C_{f i }^{d\,LR}$ involving the $A$-term by $\hat C^{d\,LR}_{fi\,A}$ and the piece containing $v_u \mu Y^{d_i}$ by $\hat C^{\prime\, d\,LR}_{fi}$. We now define 
\begin{equation}
\renewcommand{\arraystretch}{2}
\begin{array}{l}
   \hat E^d_{fi}\,=\,\dfrac{\hat C^{d\,LR}_{fi\,A}}{v_d}\,,\hspace{1.5cm} 
   \hat E^{\prime d}_{fi}\,=\,\dfrac{\hat C^{\prime\, d\,LR}_{fi}}{v_u}\,,\hspace{1.5cm} 
   \hat E^u_{fi}\,=\,\dfrac{\hat C^{u\,LR}_{fi\,A}}{v_u}\,,\hspace{1.5cm} 
   \hat E^{\prime u}_{fi}\,=\,\dfrac{\hat C^{\prime\, u\,LR}_{fi}}{v_d}\,,
   \end{array}
   \label{E-Sigma}
\end{equation}
where the parameters $\hat E_{fi}^{q}$ ($\hat E_{fi}^{\prime q}$) correspond to (non-)holomorphic Higgs-quark couplings. With these conventions, the couplings  $\epsilon^q_{ij}$ of the 2HDM in \eq{Higgs-vertices-decoupling} can be related to MSSM parameters
\begin{eqnarray}
\renewcommand{\arraystretch}{2.0}
\begin{array}{l}
 \epsilon_{fi}^{q}  = \hat E_{fi}^{\prime q}  - \left( 
 {\begin{array}{*{20}c}  
   0 & 
   {\hat E_{22}^{\prime q} \dfrac{\hat C_{12}^{q\,LR}}{m_{q_2}} } 
   & \hat E_{33}^{\prime q} \left(  \dfrac{\hat C_{13}^{q\,LR}}{m_{q_3}}  - \dfrac{\hat C _{12}^{q\;LR}}{m_{q_2}} \dfrac{\hat C_{23}^{q\,LR}}{m_{q_3}}  \right)  \\
   {\hat E_{22}^{\prime q} \dfrac{\hat C _{21}^{q\;LR}}{m_{q_2}} } 
   & 0 
   & {\hat E_{33}^{\prime q} \dfrac{\hat C _{23}^{q\;LR}}{m_{q_3}}}  \\
   {\hat E_{33}^{\prime q} \left( {\dfrac{\hat C _{31}^{q\,LR}}{m_{q_3}}  - \dfrac{\hat C _{32}^{q\;LR}}{m_{q_3}} \dfrac{\hat C _{21}^{q\,LR}}{m_{q_2}} } \right)} 
   & {\hat E_{33}^{\prime q} \dfrac{\hat C _{32}^{q\,LR}}{m_{q_3}} } 
   & 0  \\
\end{array}} \right)_{fi}  \,. 
 \end{array}
\label{Etilde}
\end{eqnarray}

\begin{figure}
\centering
\includegraphics[width=0.49\textwidth]{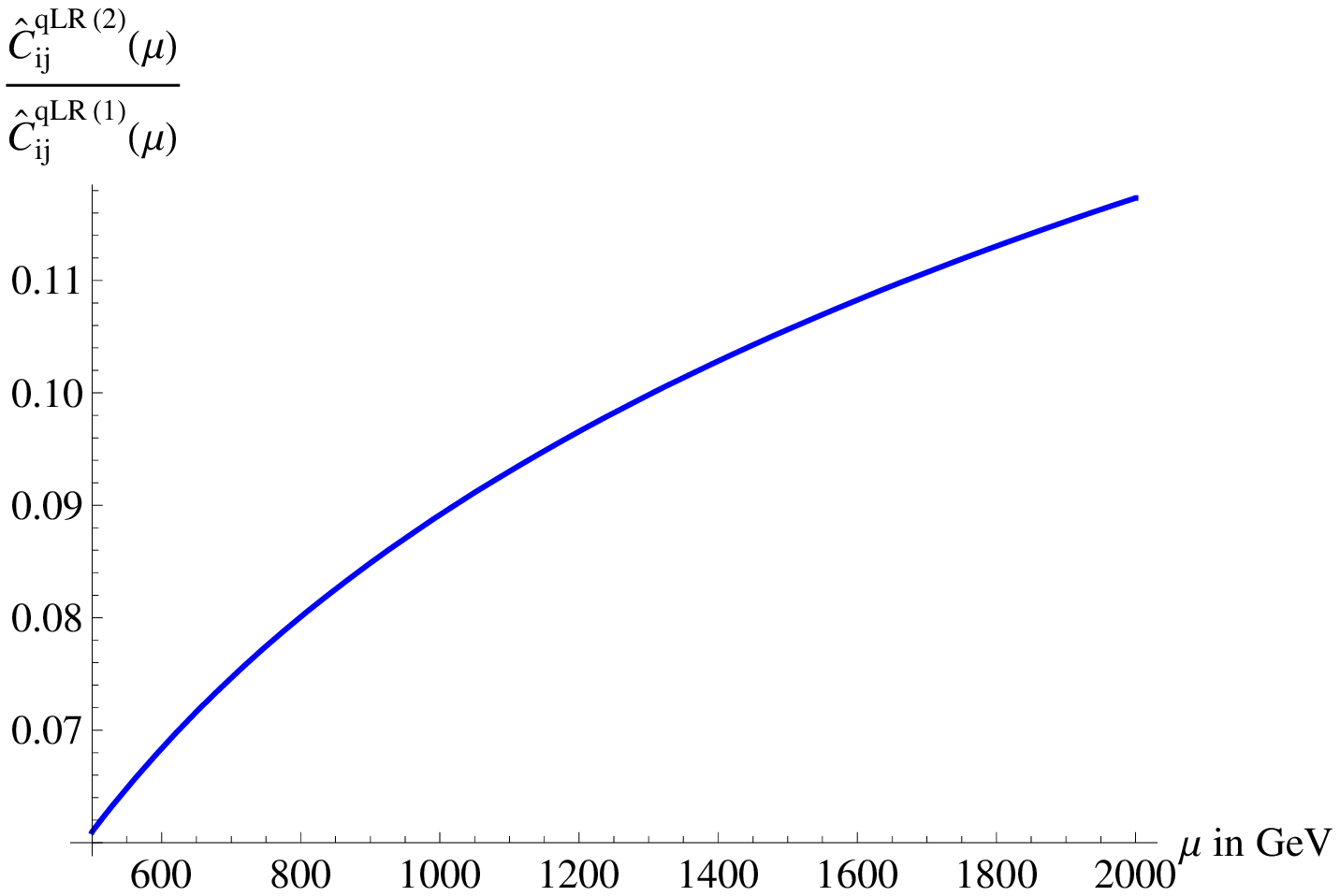}
\includegraphics[width=0.49\textwidth]{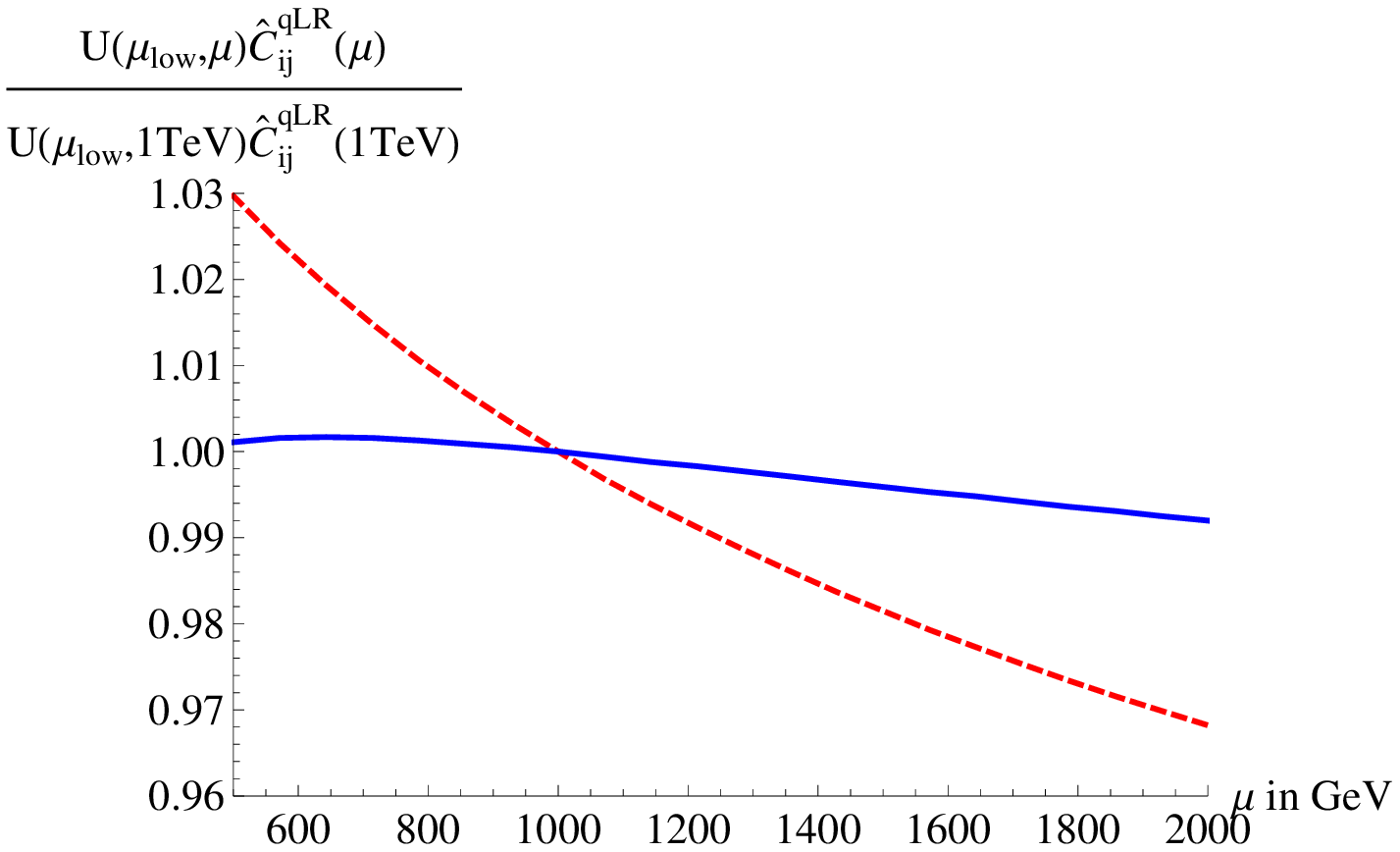}
\caption{Left: Relative importance of the two-loop corrections as a function of the matching scale $\mu$. We see that the two-loop contribution is approximately +9\% of the one-loop contribution for $\mu=M_{\rm SUSY}=1\, {\rm TeV}$.\newline
Right: Dependence on the matching scale $\mu$ of the one-loop and
  two-loop result for $\hat C _{fi}^{q\,LR}(\mu_{\rm low})$, using $M_{\rm SUSY}=1$~TeV and $\mu_{\rm low}=m_W$. Red (dashed): matching done at LO;  blue (darkest):  matching done at NLO matching.  As expected, the
  matching scale dependence is significantly reduced. For the one-loop result, $\hat  C _{fi}^{q\,LR}$ is understood to be $C_{fi}^{q\,LR\,(1)}$.
  \label{mu-abhaengigkeit}}
\end{figure}

In the matching of the MSSM on the 2HDM one can, as a byproduct, also determine the Yukawa couplings of the MSSM superpotential which is important for the study of Yukawa coupling unification in supersymmetric GUTs. Due to this importance of the chirality changing self-energies we calculated them (and thus also $\hat C^{q\,LR}_{ij}$) at the two loop-level in Ref.~\cite{Crivellin:2012zz}\footnote{The flavour conserving corrections have been calculated before in Ref.~\cite{Bednyakov:2002sf}}. The result is a reduction of the matching scale dependence (see right plot of Fig.~\ref{mu-abhaengigkeit}) while at the same time, the one-loop contributions are enhanced by a relative effect of 9\% (see left plot of Fig.~\ref{mu-abhaengigkeit}). For a numerical analysis also the LO chargino and neutralino contributions should be included by using the results of Ref.~\cite{Crivellin:2011jt}.

Concerning the tauonic $B$-decays discussed in the last section, the size of the quantities $\epsilon^u_{32,31}$ that can be generated by quantum corrections in the MSSM is by far too small to give a sizable effect even if one allows for the non-holomorphic $A^{\prime u}$.

\end{document}